\documentclass[twocolumn,english,aps,prb,superscriptaddress]{revtex4-1}
\usepackage{array}
\usepackage{amsmath}
\usepackage{graphicx}
\usepackage{amssymb}
\makeatletter
\@ifundefined{textcolor}{}
{%
 \definecolor{BLACK}{gray}{0}
 \definecolor{WHITE}{gray}{1}
 \definecolor{RED}{rgb}{1,0,0}
 \definecolor{GREEN}{rgb}{0,1,0}
 \definecolor{BLUE}{rgb}{0,0,1}
 \definecolor{CYAN}{cmyk}{1,0,0,0}
 \definecolor{MAGENTA}{cmyk}{0,1,0,0}
 \definecolor{YELLOW}{cmyk}{0,0,1,0}
}

\makeatother

\usepackage{babel}
\begin{document}

\title{A Kondo route to spin inhomogeneities in the honeycomb Kitaev
model}

\author{S. D. Das}

\affiliation{School of Physics, H. H. Wills Physics Laboratory, University of
Bristol, Tyndall Avenue, Bristol BS8 1TL, United Kingdom }

\author{K. Dhochak}

\affiliation{Department of Condensed Matter Physics, Weizmann Institute of Science, Rehovot, 76100, Israel}

\author{V. Tripathi}

\affiliation{Materials Science Division, Argonne National Laboratory, 9700 S.
Cass Ave., Lemont, IL 60439, USA}

\affiliation{Department of Theoretical Physics, Tata Institute of Fundamental
Research, Homi Bhabha Road, Navy Nagar, Mumbai 400005, India}
\begin{abstract}
Paramagnetic impurities in a quantum spin-liquid can
result in Kondo effects with highly unusual properties.
We have studied the effect of locally exchange-coupling a paramagnetic impurity
with the spin-$\frac{1}{2}$ honeycomb Kitaev model in its gapless spin-liquid phase.
The (impurity) scaling equations are found to be insensitive to the 
sign of the coupling. The weak and strong coupling fixed points are stable, with the latter
corresponding to a noninteracting vacancy and an interacting, spin-$1$ defect
for the antiferromagnetic and ferromagnetic cases respectively. The ground state in the strong coupling limit 
in both cases has a nontrivial topology associated with a finite $Z_2$ flux at the impurity site. 
For the antiferromagnetic case, this result can be obtained straightforwardly owing to the integrability of the Kitaev model with a vacancy.
The strong-coupling limit of the ferromagnetic case is however nonintegrable, and we address this 
problem through exact-diagonalization calculations with finite Kitaev fragments. 
%We find a nontrivial $Z_{2}$ flux at the impurity site in both cases. 
Our exact diagonalization calculations indicate that that the weak to
strong coupling transition and the topological phase transition occur rather close to each other and are possibly coincident.
We also find an intriguing similarity between the magnetic response of the defect and the impurity susceptibility
in the two-channel Kondo problem.
\end{abstract}
\maketitle

\section{Introduction}

A study of disorder in condensed matter systems is useful from two
perspectives. Disorder is inherent in most condensed matter systems
and often has profound effects on their properties. Incorporation
of small amounts of paramagnetic impurities in a metallic host can
result in the Kondo effect which gives the well-known logarithmic
temperature dependence of the resistivity upon cooling, and eventually
crosses over to a Fermi-liquid regime with a characteristic low energy
scale, the Kondo temperature. Conversely, impurities at low concentrations
can act as a probe providing specific signatures of the environment
they exist in. From the latter perspective, the Kondo effect is a
set of signatures of certain low-energy excitations of the host lacking
long-range magnetic order. For instance, exotic Kondo effects are known to
arise in itinerant electron magnets near 
criticality~\cite{larkin1972,varma2002,tripathi2005} and in insulating quantum spin-liquid 
systems~\cite{khaliullin97,kolezhuk06,florens06} owing to the paramagnons and spinonic excitations respectively.

A study of impurity effects in the spin-$\frac{1}{2}$ honeycomb Kitaev model~\cite{kitaev06}
is very appealing in this context. This Kitaev model
is integrable and the ground state can be either a gapless or gapped
quantum ($Z_{2}$) spin-liquid with extremely short-ranged spin correlations.~\cite{baskaran07}
The elementary excitations are not spin-$1$ bosons that one typically
expects for magnetic systems in two dimensions and higher, but emergent dispersing Majorana fermions
(spinons) and $Z_{2}$ vortices ($\pi-$flux excitations associated
with spins at the vertices of the hexagonal plaquettes) which in the
gapless phase are known to be non-Abelian anyons.~\cite{kitaev06} 
Experimental realization looks increasingly imminent with several interesting proposals to realize Kitaev physics
in two-dimensional quantum-compass materials such as the alkali iridates~\cite{jackeli09} and ruthenium trichloride,~\cite{banerjee} 
and independently in cold-atom optical lattices.~\cite{duan03} 

Introducing a paramagnetic impurity into the model through local exchange-coupling of the impurity spin with a host (Kitaev) spin 
results in a highly unusual Kondo effect~\cite{dhochak2010} owing to the peculiar elementary excitations in
the host. For an $S=1/2$ Kitaev model with an energy scale $J$ coupled locally to a spin-$S$ impurity, 
the perturbative scaling equations for the impurity coupling
$K$ turn out to be independent of its sign, with an intermediate coupling unstable 
fixed point $|K|\sim J/S$ separating weak and strong coupling regimes.
Such scaling differs qualitatively from the Kondo effect in metals (and graphene,~\cite{withoff90}) 
where a nontrivial effect is seen only for antiferromagnetic coupling, but is similar to the Kondo 
scaling reported for paramagnetic impurities in certain pseudogapped bosonic spin-liquids.~\cite{florens06} 
The distinguishing feature of the Kitaev-Kondo problem is that the weak and strong coupling limits 
correspond to different topologies of the ground state.~\cite{dhochak2010}

Despite the insensitivity of the scaling equations to the sign of impurity coupling, 
the strong coupling limits for $K>0$ and
$K<0$ are very different physically. In the antiferromagnetic case
($K>0$), the strong-coupling limit for an $S=1/2$ impurity spin
corresponds to a spin singlet at the impurity site - equivalent to
the Kitaev model with a missing site, which is an integrable model. In the ferromagnetic case, the
strong-coupling limit corresponds to a non-integrable problem where one of
the sites has $S=1,$ while the rest have $S=1/2.$ 
%For any sign or strength of impurity coupling, one can construct three
%new conserved quantities~\cite{dhochak2010}: composite operators consisting of the product of an impurity spin
%component and two of the three flux operators enclosing the defect
%(a feature reminiscent of composite fermions in quantum Hall literature).
%which, remarkably, satisfy an SU(2) algebra. This immediately leads
%to an important conclusion for the strong coupling limit - the ground
%state of the Kitaev model has a two-fold degeneracy when a site is
%removed or when one of the sites is a spin-$1$ defect. For weak-coupling, the two-fold degeneracy
%of the ground state is nothing special and corresponds to the degeneracy of the isolated
%impurity spin.

The problem of missing sites (spinless vacancies) in the Kitaev model
has received much attention in recent times. It was independently
reported in Ref. \onlinecite{dhochak2010} and Ref. \onlinecite{willans10} that the ground state of
the Kitaev model with a missing site is associated with a finite $Z_{2}$
flux through the defect. That would not be the case, for example in
graphene, where although the Dirac fermions have the same dispersion
as the emergent Majorana fermions in the Kitaev model, the phases
of the intersite hopping matrix elements in graphene are identical
for every bond and do not change upon the creation of defects. In
contrast, the phases of the intersite hopping elements of the Majorana
fermions in the Kitaev model are a degree of freedom and can take
values $0$ or $\pi.$ For the missing site problem, the magnetic
susceptibility is predicted\cite{willans10,willans2011} to have logarithmic singularities
both as a function of the magnetic field as well as the temperature.
%At the same time, one can explicitly demonstrate the presence of an
%odd number of bound, zero energy Majorana fermions at the vacancy
%site\cite{dhochak2010}. 
Some of the singularities in magnetic susceptibility
reported in Refs. \onlinecite{willans10,willans2011} are reminiscent of the two-channel
Kondo problem, and we shall later discuss a connection between such
singularities and the presence of bound, zero energy Majorana fermions
in the Kitaev model with a vacancy~\cite{dhochak2010} as well as in the two-channel Kondo model.~\cite{coleman1995} Vacancy induced
spin textures have also been studied by exact diagonalization\cite{trousselet2011} 
of finite clusters of up to 24 spins described by more
general (and nonintegrable) Kitaev-Heisenberg models in the presence
of a small magnetic field. In Ref. \onlinecite{trousselet2011}, it was reported that
a vacancy induces longer ranged spin-spin correlations that extend
beyond the single bond correlations one has in the defect-free Kitaev
model.~\cite{baskaran07} The low-energy properties of the Kitaev model with a 
random and dilute concentration of vacancies are also quite interesting. Such rare but 
locally singular perturbations are predicted to result in qualitatively different low-energy properties
compared to that expected for Gaussian white noise type disorder.\cite{willans2011} 

In contrast to the understanding we currently have on the effects
of single and multiple vacancies in the Kitaev model, much less is
known about the effect of spinful defects where the defect site has
a different spin from the host sites. Part of the reason is that while
the vacancy problem is integrable and affords a simplification of
a difficult problem where interactions and disorder are both present,
the problem with an $S=1$ defect cannot be reduced to a noninteracting
one. Some progress was made in Ref. \onlinecite{dhochak2010} where it was explicitly
demostrated that the ground states of both the vacancy as well as
$S=1$ defect problems have a two-fold degeneracy. However, it could
not be established whether the $S=1$ defect was also associated with
a finite $Z_{2}$ flux that is the case when a vacancy is present.
It was also not demonstrated whether the strong coupling fixed points
were indeed stable. The stability is an important issue, for otherwise
one would expect new intermediate coupling stable fixed points and
not only would our understanding of the Kondo effect in the Kitaev
model be incomplete, but also the paramagnetic impurity route for
generating vacancies and spinful defects would no longer be appropriate.
The latter issue is of interest from a practical point of view too
since it would make it possible to generate nonabelian anyons conveniently
using a spin-polarized STM tip to bind a Kitaev spin ferromagnetically
or antiferromagnetically. It is also not clear whether the topological transition
and the magnetic transitions are coincident or occur at the same value of 
the impurity coupling strength.

In this paper we address these open questions and make the following new findings. 
We demonstrate the stability of the strong coupling
limit demonstrated for the ferromagnetic and antiferromagnetic cases which implies stability of the
spin-$0$ vacancies and spin-$1$ defects created through this route.  
We perform exact diagonalization calculations for a finite fragment of the Kitaev model coupled to
a paramagnetic impurity and show that while for weak impurity coupling,
the ground state corresponds to zero $Z_{2}$ flux at the impurity
site, for strong coupling, the ground state has a finite flux irrespective
of the sign of impurity coupling. For the value of impurity coupling
that corresponds to this topological transition, we also observe the
total spin at the impurity site going to zero or one depending on
the sign of coupling - this establishes that the the topological as
well as the weak coupling to strong coupling transitions occur very
close to each other and possibly at the same point.  As
a corollary, we find that the ground state with a spin-$1$ defect corresponds
to a finite flux at the defect site. Finally, we report an intriguing 
connection between the susceptibilities of a vacancy in the Kitaev model 
and of the magnetic impurity in a two-channel Kondo model.

The rest of the paper is organized as follows. Sec.\ref{sec:model} provides a brief introduction
to the honeycomb spin-$\frac{1}{2}$ Kitaev model. In Sec.\ref{sec:Kondo} we study the effect of
coupling an external paramagnetic impurity to the Kitaev host through a local exchange coupling.
A poor man's scaling analysis of the impurity coupling reveals an unstable fixed point separating
the weak and strong-coupling regimes. The stability if the strong coupling fixed point
is demonstrated for both ferromagnetic and antiferromagnetic impurity couplings. It is also shown that the 
strong and weak coupling limits correspond to different topology of the ground state. New conserved quantities
are identified which are composite operators of an impurity spin component and two Kitaev flux operators.
Sec.\ref{sec:numerical} contains the result of exact diagonalization calculations of finite Kitaev fragments
coupled to external spins. The key findings in this section are (a)establishing that a finite $Z_2$ flux is associated
with the ground state of the spin$-\frac{1}{2}$ Kitaev model with a spin$-1$ defect - just as in the case of a vacancy,
and (b) the magnetic (Kondo) and topological transitions occur very close to each other and are possibly coincident.
In Sec.\ref{sec:two-channel} we discuss the intriguing parallels between the low-temperature 
magnetic response of the Kitaev model with a missing site and the two-channel Kondo model. Sec.\ref{sec:discussion} contains
a discussion of the results and possible future directions.

\section{The spin-$1/2$ Kitaev model on the honeycomb lattice}\label{sec:model}
\begin{figure}
\begin{centering}
\includegraphics[width=0.9\columnwidth]{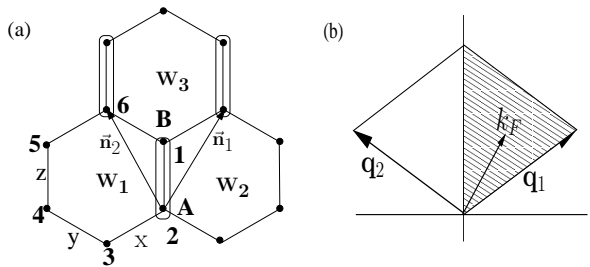} 
\par\end{centering}
\caption{\label{fig:basis}(a) Schematic of a fragment of the Kitaev lattice showing the
$A$ and $B$ sites and the $x,$ $y$ and $z$ types of bonds. (b)
Figure showing the reciprocal lattice vectors for the $A$ sublattice.
The Dirac point for the massless Majorana fermions is denoted by $k_{F}$
and momentum summations are over the (shaded) half Brillouin zone.}
\end{figure}

Kitaev's spin-$\frac{1}{2}$ honeycomb lattice model for a quantum spin liquid 
is a model of direction dependent nearest neighbour exhchage interactions on a 
honeycomb lattice.~\cite{kitaev06} The Hamiltonian for this model is given by
\begin{equation}
H_{0}=-J_{x}\!\!\!\sum_{\text{\ensuremath{x}-links}}\sigma_{j}^{x}\sigma_{k}^{x}
-J_{y}\!\!\!\sum_{\text{\ensuremath{y}-links}}\sigma_{j}^{y}\sigma_{k}^{y}-J_{z}
\!\!\!\sum_{\text{\ensuremath{z}-links}}\sigma_{j}^{z}\sigma_{k}^{z},\label{
eq:Hamiltonian}
\end{equation}
where the three bonds at each site (see Fig.\ref{fig:basis}) are labeled as $x$, $y$ and $z.$ 
The model is exactly solavable.~\cite{kitaev06}  As was shown by Kitaev, the flux 
operators
$W_{p}=\sigma_{1}^{x}\sigma_{2}^{y}\sigma_{3}^{z}\sigma_{4}^{x}\sigma_{5}^{y}
\sigma_{6}^{z}$
defined for each elementary plaquette $p$ are conserved (Fig.\ref{fig:basis}),
with eigenvalues $\pm1,$ and form a set of commuting observables. The 
Kitaev spins can be represented in terms of Majorana fermions
$b_{i}^{x},\, b_{i}^{y},\, b_{i}^{z},\, c_{i}$ as 
$\sigma_{i}^{\alpha}=ib_{i}^{\alpha}c_{i}.$ 
This representation spans a larger Fock space, and we restrict to the physical 
Hilbert space of the spins by choosing the gauge\cite{kitaev06} 
$D_{i}=ib_{i}^{x}b_{i}^{y}b_{i}^{z}c_{i}=1.$

For each $\alpha-$type bond, $u_{ij}^{\alpha}=ib_{i}^{\alpha}b_{j}^{\alpha}$
is also conserved and the flux operators can be written as a product of 
$u_{ij}$'s on the plaquette $\Pi_{\langle ij\rangle\in \text{Plaq.}}u_{ij} .$
The ground state manifold corresponds to a vortex-free state where all $W_{i}$ are equal. 
In the vortex-free state,
we can fix all $u_{ij}=1$ (corresponds to $W_{p}=1$) and the Hamiltonian
can be written as a tight-binding model of noninteracting Majorana fermions. The
reduced Hamiltonian for this ground state manifold is given by 
$H_{0}=\frac{i}{4}\sum_{jk}A_{jk}c_{j}c_{k},$
where $A_{jk}=2J_{{\alpha}_{_{jk}}}$ if $j,\, k$ are neighboring
sites on an $\alpha-$bond and zero otherwise. The excited states
(with finite vorticity) are separated from the ground state manifolds
by a gap of order $J_{\alpha}.$ 

The free Majorana fermion hopping Hamiltonian can be diagonalized in momentum 
space by defining the Bravais lattice with
a two-point basis (Fig.\ref{fig:basis}). In momentum space,
\begin{align}
 H_{0}=&\frac{1}{4}\sum_{\mathbf{q}>0,\alpha}\epsilon_{\alpha}(\mathbf{q})a_{
\mathbf{q},\alpha}^{\dagger}a_{\mathbf{q},\alpha},
\end{align}
with $\epsilon_{\alpha}(\mathbf{q})=\pm|f(\mathbf{q})|,\:\: 
f(\mathbf{q})=2 i (J_{x}e^{ia_0\mathbf{q\cdot n_{1}}}+J_{y}e^{ia_0\mathbf{q\cdot 
n_{2}}}+J_{z})$, where $a_0$ is the nearest neighbor spin distance. The eigenstates 
are 
$a_{\mathbf{q},0}=\tilde{c}_{\mathbf{q},A}+\tilde{c}_{\mathbf{q},B}e^{-i\tilde{
\alpha}(\mathbf{q})}$
and 
$a_{\mathbf{q},1}=\tilde{c}_{\mathbf{q},A}-\tilde{c}_{\mathbf{q},B}e^{-i\tilde{
\alpha}(\mathbf{q})},$ with $\alpha(q)$ being the phase of $f(q).$

The Kitaev model has gapless excitations for a region of parameter space
where $J_\alpha$'s satisfy the triangle inequalities $|J_x| +|J_y| \geq |J_z|$ etc. 
and a gapped spectrum outside this parameter regime. The gapless phase has a point Fermi 
surface where $\epsilon(\mathbf{k}_{F})=0$ and
$\epsilon(\mathbf{q})$ has a linear dispersion around $\mathbf{k}_{F}$
(Fig.\ref{fig:basis}).
%The position of the Fermi point in the Brillouin
%zone depends on the relative values of $J_{x},$ $J_{y}$ and $J_{z}.$
For simplicity, 
%and without loss of generality, 
we will assume
$J_{x}=J_{y}=J_{z}=J$ for further analysis. For this case, the Fermi points are at 
$(\pm 4\pi/3\sqrt{3}a_0,0).$

The ground state of the Kitaev model is a quantum spin liquid with only nearest 
neighbor spin-spin correlations.~\cite{baskaran07} On an $\alpha-$bond, only $\langle
\sigma_i^{\alpha}\sigma_j^{\alpha}\rangle$ is non zero and other two spin
correlations are zero. Four spin bond-bond correlations are long-ranged with power-law 
decay in the gapless phase of the Kitaev model.

\section{Topological Kondo effect}\label{sec:Kondo}
Consider a spin $S$ impurity locally exchange-coupled to a host (Kitaev) spin at an $A$
site ($\mathbf{r}=0$): 
\begin{align}
V_{K}= & i\sum_{\alpha}K^{\alpha}S^{\alpha}b^{\alpha}c_{A}=i\sum_{\mathbf{q}\in
HBZ,\alpha}\frac{K^{\alpha}}{\sqrt{2N}}S^{\alpha}b^{\alpha}(\tilde{c}_{\mathbf{q},A}
+\tilde{c}_{\mathbf{q},A}^{\dagger})\nonumber \\
\equiv & \frac{1}{\sqrt{N}}\sum_{\mathbf{q}\in
HBZ,\alpha,\beta}Q^{\alpha}S^{\alpha}b^{\alpha}(a_{\mathbf{q},\beta}+a_{\mathbf{q},
\beta}^{\dagger}).\label{eq:Kondo-term}
\end{align}

\begin{figure}
\begin{centering}
\includegraphics[width=0.8\columnwidth]{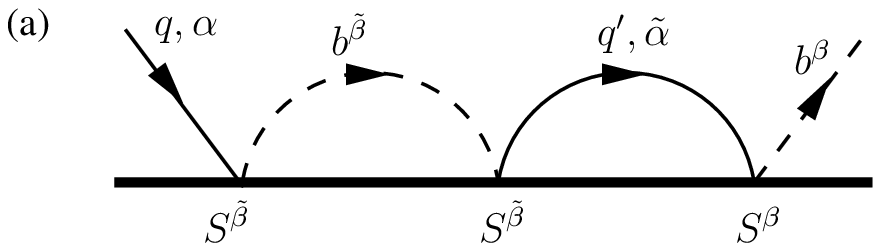}\\
 \vspace{8mm}
 \includegraphics[width=0.8\columnwidth]{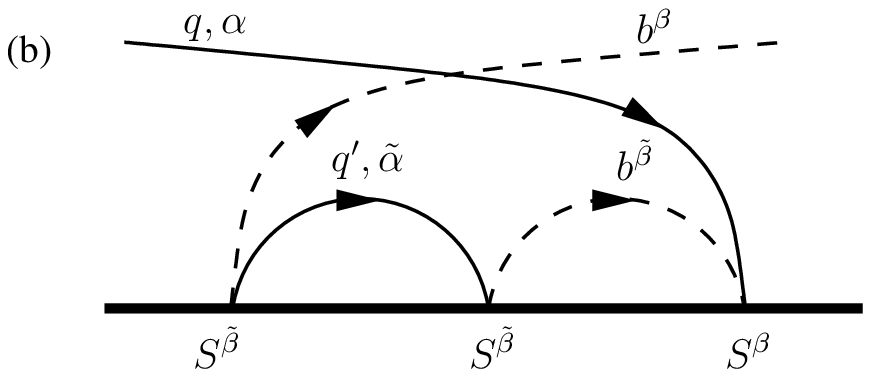}\\
\par\end{centering}
\caption{Diagrams contributing to the scaling of Kondo coupling $K^{\alpha}$.}
\centering{}\label{fig:Kondo-on-site} 
\end{figure}

We perform a poor man's scaling analysis\cite{anderson70,hewson92} for the Kondo coupling 
$K$ to study the screening of the impurity spin by the host excitations. To
study the system properties at low temperatures, we can compute the effective Hamiltonian for a
reduced bandwidth for the fermionic excitations ($-D+\delta D \text{ to } D-\delta D$) 
by integrating out the excitations in the band edges (($-D \text{ to } -D+\delta D) 
\text{ and } (D \text{ to } D-\delta D)$). This process is successively repeated to 
get a scaling law for the coupling constants in the Hamiltonian. We consider the 
Lippmann-Schwinger expansion for the $T-$matrix element,
$\langle\Omega,b^{\beta}|K^{\beta}S^{\beta}b^{\beta}c_{a,A}|\Omega+(\mathbf{q},
\alpha)\rangle.$ 
making a perturbation expansion $T=T^{(1)}+T^{(2)}+\cdots$ in increasing
powers of $K$ and following its variation as a function of the decrease
of the bandwidth $(-D,D),$ we find that the first correction to the bare $T-$matrix comes
from two \emph{third} order terms (see Fig. \ref{fig:Kondo-on-site}).
The contribution from on-site scattering (Fig.\ref{fig:Kondo-on-site}a) is 
\begin{align}
T^{(3),a} &
=\langle\Omega,b^{\beta}|V_{K}G_{0}^{+}(E)V_{{K}}G_{0}^{+}(E)V_{K}|\Omega+(\mathbf{q},
\alpha)\rangle\nonumber \\
\nonumber \\
&
\!\!\!\!\!\!\!\!\!\!\!\!\!\!\!=\frac{Q^{\beta}S^{\beta}}{N}
\!\!\!\!\!\!\!\!\!\sum_{(D-\delta
D)\le|\epsilon_{q'}|,|\epsilon_{q''}|\le
D,\tilde{\alpha},\tilde{\beta},\tilde{\alpha}'}
\!\!\!\!\!\!\!\!\!\!\!\!\!\!\!\!\!\!\!\!\!(Q^{
\tilde{\beta}})^{2}(S^{\tilde{
\beta}})^{2}\langle b^{\beta}|\:
b_{\beta}^{\dagger}(a_{\mathbf{q}'',\tilde{\alpha}'}+a_{\mathbf{q}'',\tilde{\alpha}'}^
{\dagger})\nonumber \\
 &
\hspace{0.5cm}\times  G_{0}^{+}(\epsilon)\: 
b_{\tilde{\beta}}(a_{\mathbf{q}',\tilde{\alpha}}+a_{\mathbf{q}',\tilde{\alpha}}^{
\dagger})\: G_{0}^{+}(\epsilon)\:
b_{\tilde{\beta}}^{\dagger}c_{\mathbf{q},\alpha}|(\mathbf{q},\alpha)\rangle\nonumber
 \\
\nonumber \\
 &
\!\!\!\!\!\!\!\!\!\!\!\!\!\!\!\!\!\!\!\!=-\frac{Q^{\beta}S^{\beta}}{N}\!\sum_{
\mathbf{q}', \tilde{
\beta}}(Q^{\tilde{\beta}})^{2}(S^{\tilde{\beta}})^{2}\left\langle
a_{\mathbf{q}',1}^{\dagger}a_{\mathbf{q}',1}\frac{1}{E-(H_{0}-\epsilon_{q',1})}
\right. \nonumber \\
& \hspace{1.8cm} \left. +a_{
\mathbf{q}',0}a_{\mathbf{q}',0}^{\dagger}\frac{1}{E-(H_{0}+\epsilon_{q',0})}
\right\rangle\! \frac{1}{E-\epsilon_{b}}\:\nonumber \\
\nonumber \\
 & \!\!\!\!\!\!\!\!\!\!\!\!\!\!\!\simeq-2Q^{\beta}S^{\beta}\frac{\rho(D)a^{2}|\delta
D|}{E-D}\cdot\frac{1}{E-J}\sum_{\tilde{\beta}}(Q^{\tilde{\beta}})^{2}(S^{\tilde{\beta}
})^{2}. \label{eq:T3a}
\end{align}
\\
Here $\rho(D)$ is the density of states at the band edge, $a$ is
the lattice constant and $G_{0}(E)=(E-H_{0}+i\delta)^{-1}.$ 

Similarly,
the contribution from Fig. \ref{fig:Kondo-on-site}(b) is 
\begin{align}
T^{(3),b}&=\frac{Q^{\beta}S^{\beta}}{N}
\!\!\!\!\!\!\!\!\!\!\!\!\!\!\!\!\!\!\!\!\!\!\!\!\!\!\!\sum_{(D-\delta
D)\le|\epsilon_{q'}|,|\epsilon_{q''}|\le
D,\tilde{\alpha},\tilde{\beta},\tilde{\alpha}'}
\!\!\!\!\!\!\!\!\!\!\!\!\!\!\!\!\!\!\!\!\!\!\!\!\!\!\!(Q^{
\tilde{\beta}})^{2}(S^{\tilde{
\beta}})^{2}\langle b^{\beta}|\:
b_{\tilde{\beta}}^{\dagger}(a_{\mathbf{q},{\alpha}}+a_{\mathbf{q},{
\alpha }}^{\dagger})\nonumber \\
& \hspace{-0.8cm}\times  G_{0}^{+}(\epsilon)\: 
b_{\tilde{\beta}}(a_{\mathbf{q}',\tilde{\alpha}}+a_{\mathbf{q}',\tilde{\alpha}}^{
\dagger})\: G_{0}^{+}(\epsilon)\:
b_{{\beta}}^{\dagger}(a_{\mathbf{q}',\tilde{\alpha}}+a_{\mathbf{q}',\tilde{\alpha}}^{
\dagger})|(\mathbf{q},\alpha)\rangle\nonumber
 \\
\nonumber \\ 
& \hspace{-0.4cm}
=-\frac{Q^{\beta}S^{\beta}}{N}
\!\!\!\!\!\!\!\!\!\!\!\!\!\!\!\!\!\!\!\!\!\!\!\!\!\sum_{(D-\delta
D)\le|\epsilon_{q'}|,|\epsilon_{q''}|\le
D,\tilde{\alpha},\tilde{\beta},\tilde{\alpha}'}
\!\!\!\!\!\!\!\!\!\!\!\!\!\!\!\!\!\!\!\!\!\!\!\!(Q^ {
\tilde{\beta}})^{2}(S^{\tilde{
\beta}})^{2} \:\langle\Omega,b^{\beta} | \:b_{\tilde{\beta}}^{\dagger} 
c_{\mathbf{q},\alpha}  
b^{\dagger}_{\beta} c^{\dagger}_{\mathbf{q},\alpha}\: \nonumber \\
& \hspace{-0.1cm} \times \frac{1}{E-(H_{0}+ \epsilon_{b}
+\epsilon_{q,0})} \: b_{\tilde{\beta}} 
(a_{\mathbf{q}',\tilde{\alpha}}+a^{\dagger}_{\mathbf{q}',\tilde{\alpha}}
)\nonumber 
\\
& \hspace{-0.1cm} \times \frac{1}{
E-(H_{0} +\epsilon_ {b} + \epsilon_{q,0})} \:
(a_{\mathbf{q}',\tilde{\alpha}}+a^{\dagger}_{\mathbf{q}',\tilde{\alpha}}) |
\Omega\rangle 
\nonumber \\
 \nonumber \\
 = & - \frac{Q^{\beta}S^{\beta}}{N} \sum_{\mathbf{q}',\tilde{\beta}}
(Q^{\tilde{\beta}})^2
(S^{\tilde{\beta}})^2 \:\frac{1}{E-2 \epsilon_{b}} \: 
\nonumber \\
& \hspace{0.4cm} \times  \langle a^{\dagger}_{\mathbf{q}',1} a_{\mathbf{q}',1}
\frac{1}{E-(-\epsilon_{q',1}+\epsilon_{b}+\epsilon_{q,0} )}\nonumber \\
& \hspace{0.4cm} + a_{\mathbf{q}',0}
a^{\dagger}_{\mathbf{q}',0}\frac{1}{E-(\epsilon_{q',0}+\epsilon_{b}+\epsilon_{q,0} 
)} \rangle
\nonumber \\  
\nonumber \\
 & \hspace{-0.4cm}\simeq-2Q^{\beta}S^{\beta}\:\frac{\rho(D)a^{2}|\delta
D|}{E-D-J}\cdot\frac{1}{E-2J}\sum_{\tilde{\beta}}(Q^{\tilde{\beta}})^{2}(S^{\tilde{
\beta}})^{2}.\label{eq:T3b}
\end{align}

Adding the two contributions (taking $E\simeq0$), 

\begin{align}
T^{(3)} & \simeq2Q^{\beta}S^{\beta}\rho(D)\frac{a^{2}\delta
D}{\epsilon_{b}}\sum_{\tilde{\beta}}(Q^{\tilde{\beta}})^{2}(S^{\tilde{\beta}})^{2}
\left\lbrace \frac{1}{D}+\frac{1}{2(D+J)}\right\rbrace .\label{eq:T3}
\end{align}
 Here we have taken $E,\epsilon_{q,\alpha}\ll D,\, J$ and neglected
them.

If either the impurity is a $S=\frac{1}{2}$ spin, or the Kondo interaction
is rotationally symmetric, the above contribution renormalizes the
Kondo coupling constant. However for $S\neq\frac{1}{2}$ with anisotropic
coupling, new terms are generated and one needs to go to higher order
diagrams to obtain the scaling of these new coupling terms. For $S=1/2$
or for symmetric impurity coupling we thus have \begin{align}
\delta K & \sim-2K^{3}S(S+1)\rho(D)a^{2}\frac{\delta D}{J}\left\lbrace
\frac{1}{D}+\frac{1}{2(D+J)}\right\rbrace .\label{eq:delta-K}\end{align}
 Just as for the Kondo effect in graphene\cite{withoff90}, owing
to the change in the density of states with bandwidth (here $\rho(\epsilon)=(1/2\pi
v_{F}^{2})|\epsilon|\equiv C|\epsilon|$),
we also need to consider the change in $K$ due to the rescaling done in order to keep
the total number of states fixed. This gives a contribution $K\rightarrow
K(D'/D),\quad(D'=D-|\delta D|).$
In addition, as we shall scale the bandwidth $D$ to smaller values,
the second term in Eq. \ref{eq:delta-K} may be dropped.
Thus 
\begin{align}
\delta K & \simeq-2K^{3}S(S+1)\rho(D)a^{2}\frac{\delta D}{DJ}+K\frac{\delta
D}{D}\nonumber \\
 & =-K\frac{\delta D}{D}\left(2K^{2}a^{2}CDS(S+1)/J-1\right).\label{eq:iso-kondo}
\end{align}
 Thus, as we decrease the bandwidth by integrating out the high energy
excitations, the effective coupling $K$ has an unstable fixed point
at $K_{c}=\sqrt{J/[2a^{2}\rho(D)S(S+1)]};$ or in other words,
$K_{c}\sim\sqrt{J/S^{2}a^{2}CD}\sim J/S.$
Here we used $D\lesssim J$ and $C\sim1/(Ja)^{2}.$ Clearly for $K>K_{c},$
the coupling flows to infinity independent of the nature of coupling
(ferromagnetic or antiferromagnetic), while for $K<K_{c},$ the coupling
flows to zero. For anisotropic Kondo coupling we can show
\begin{align}
\delta K_{z,\pm} & \sim-K_{z,\pm}\frac{\delta
D}{D}\left[2a^{2}\rho(D)S(S+1)\frac{K_{z}^{2}+K_{+}K_{-}}{J}-1\right].\label{
eq:anis-Kondo}
\end{align}
The two-parameter Kondo flow is therefore given by 
\begin{align}
\frac{\delta K_{z}}{\delta
K_{\pm}}=\frac{K_{z}}{K_{\pm}}\Rightarrow\frac{K_{z}}{K_{\pm}}=\text{const}.\label{
eq:2-param-scaling}
\end{align}

A comparison of the Kondo effect in graphene
\cite{withoff90},
a bosonic spin bath \cite{florens06} and the Kitaev model are shown
in Table \ref{tab:comparison}.

\begin{table*}
\begin{small}
\hspace{-5mm}\begin{tabular}{|>{\raggedright}p{0.45in}|>{\raggedright}p{1.55in}
|>{
\raggedright } p {
1.8in}|>{\raggedright}p{1.55in}|}
\hline 
 & Graphene  & $Z_{2}$ bosonic spin bath with pseudogap density of states
$\rho(\epsilon)=C|\epsilon|.$  & Kitaev, honeycomb lattice\tabularnewline
\hline
\hline 
Kondo scaling  & Unstable intermediate coupling fixed pt. only for AFM coupling.
Only
AFM flows to strong coupling above unstable fixed pt.  & Flow direction is
independent
of the sign of magnetic impurity coupling.
Unstable intermediate coupling fixed pt. for both FM and AFM.  & Scaling same as
$Z_{2}$ bosonic spin bath case. However a topological
transition is associated with the unstable fixed point.\tabularnewline
\hline
\end{tabular}
\caption{\label{tab:comparison}Comparison of Kondo effect 
in graphene, a $Z_{2}$ bosonic spin bath with a pseudogap density
of states and the Kitaev model on the honeycomb lattice.}
\end{small}
\end{table*}

\subsection{Stability of strong coupling point}
The poor man's scaling analysis is only valid for small Kondo couplings as the
perturbation theory breaks down much before the critical value of the coupling. While we have shown that 
the coupling flows to larger values above the
critical value $K_c,$ it remains to be seen whether there is any other fixed point beyond $K_c$ but less than
the $\infty.$ Below we study the model in the strong coupling limit and see if it is a stable fixed point.
In the strong coupling limit, $K$ is the largest energy scale and the impurity spins
forms a singlet/triplet with the Kitaev spin at origin.

We consider the Hamiltonian such that the Kondo term and the Kitaev model with
one spin
missing ($H_{K-}$) constitute the unperturbed Hamiltonian and Kitaev coupling to the
site at origin is the perturbation:
\begin{align}
 H_0 & = K \mathbf{S}\cdot \mathbf{\sigma}_0 + H_{K-},\\
V & = J (\sigma_0^x \sigma_1^x+\sigma_0^y \sigma_2^y + \sigma_0^z \sigma_3^z).
\end{align}
For antiferromagnetic Kondo coupling ($K>0$), the ground state consists of a Kondo
singlet of $S \text{ and }\sigma_0$ and the Kitaev model with one spin missing. The
perturbation term causes transitions from singlet to triplet states of the Kondo
singlet. We use effective Hamiltonian scheme\cite{primas63} to include the
effects of the
perturbation terms within the projected ground state subspace.
\begin{align}
H_{\text{eff}} = \text{e}^{iQ} H \text{e}^{-iQ},
\end{align}
where $Q$ is chosen such that the terms which take us out of the reduced Hilbert
space are canceled order by order.
This gives the reduced Hamiltonian as
\begin{align}
& H_{\text{eff}}  = H_0 + H_1 + H_2 + O(V^3),\\
&\langle \alpha|H_1| \beta \rangle   = \langle \alpha|V| \beta \rangle, \\
&\hspace{-1mm}\langle \alpha|H_2| \beta \rangle   =\frac{1}{2} \!\!\! 
\sum_{\gamma \neq \alpha, \beta}\!\!\!
\langle \alpha|V|\gamma\rangle \langle \gamma|V| \beta \rangle \!\left(
\frac{1}{E_{\alpha}-E_{\gamma}}+\frac{1}{E_{\beta}-E_{\gamma}}\!\right)\!,
\end{align}
where $\alpha, \beta$ belong to the ground state manifold and $\gamma$ belongs to
excited state manifold. 
The eigenstates of the Kondo term are singlet $|s\rangle$ and triplet states
$|t,(0,\pm1)\rangle$:
\begin{align}
& |s\rangle = \frac{1}{\sqrt{2}}\left( |\uparrow,\Downarrow\rangle - |\downarrow,
\Uparrow \rangle \right),\\
& |t,1\rangle = |\uparrow,\Uparrow\rangle,\\
& |t,0\rangle = \frac{1}{\sqrt{2}}\left( |\uparrow,\Downarrow\rangle + |\downarrow,
\Uparrow \rangle \right),\\
& |t,-1\rangle = |\downarrow,\Downarrow\rangle.
\end{align}
Here $\uparrow$ refers to the Kitaev spin and $\Uparrow $ refers to the
impurity spin
state.

\subsubsection*{Antiferromagnetic Kondo coupling}
For the antiferromagnetic Kondo coupling case, ground state is the singlet
state. As  $\langle s |V|s \rangle = 0$, $H_1=0$ and
\begin{align}
 & \langle s,K_-|H_2|s,K_-'\rangle = \frac{1}{2}\sum_{t, K_-''} 
\langle
s,K_-|V|t,K_-''\rangle\nonumber \\
 & \hspace{10mm}\times \langle t,K_-''|V| s,K_-'\rangle \left(
\frac{1}{E_0-E_{t}}+\frac{1}{E_{0}'-E_{t}}\right).
\end{align}
Here, $K_-$ denotes the eigenstates of the Kitaev model with the spin at origin
missing. Since
change in energy of the Kitaev state is $\sim J \ll K$, we ignore their contribution
in the energy denominators of the perturbation term. The matrix elements of $H_2$ are then
\begin{align}
 &\langle s,K_-|H_2|s,K_-'\rangle \\
 & \simeq \frac{J^2}{E_0-E_t}\!\!\!\sum_{t,
K_-'',\alpha,\beta}\!\!\!\!\!\! \langle
s|\sigma_0^{\alpha}|t\rangle \langle t|\sigma_0^{\beta}|
s\rangle \langle K_-|\sigma_{\alpha}^{\alpha}|K_-''\rangle \langle
K_-''|\sigma_{\beta}^{\beta}| K_-'\rangle \\
& \simeq -\frac{J^2}{K}\sum_{\alpha,\beta}
\langle s|\sigma_0^{\alpha}(\mathbf{1}-|s\rangle \langle s|)\sigma_0^{\beta}|s\rangle
\langle K_-|\sigma_{\alpha}^{\alpha}\sigma_{\beta}^{\beta}| K_-'\rangle \\
& = -\frac{3 J^2}{K}\sum_{\alpha,\beta} \langle
K_-|\sigma_{\alpha}^{\alpha}\sigma_{\beta}^{\beta}| K_-'\rangle.
\end{align}
%
%\begin{align}
%& H_2  \simeq -\frac{3 J^2}{K}
%\sum_{\alpha,\beta} \sigma_{\alpha}^{\alpha}\sigma_{\beta}^{\beta} 
%\end{align}
Here, in $\sigma_{\alpha}^{\alpha},$ the subscript $\alpha$ refers to a neighoring site
of the origin in the direction of the $\alpha-$bond.
Thus, in the antiferromagnetic coupling case, the Kondo singlet decouples from the
rest of the Kitaev model and a small interaction ($ J^2/K \ll J $) is generated
between the Kitaev spin at the origin and the sins at the three neighboring sites 
in the second order perturbation. The
strong coupling fixed point is thus a stable fixed point and is equivalent to the Kitaev
model with one site missing.

\subsubsection*{Ferromagnetic Kondo coupling}
In the ferromagnetic Kondo coupling case, the triplet states form the ground state
manifold. We perform degenerate perturbation theory to get the effective Hamiltonian:
\begin{align}
 \langle t',K_-'|H_1| t, K_-\rangle  & = J \langle t',K_-'|V| t,K_- \rangle \\
& =  \sum_{\alpha} \langle t'|\sigma_0^{\alpha}| t\rangle \langle
K_-'|\sigma_{\alpha}^{\alpha}|K_- \rangle.
\end{align}
If we calculate the matrix elements of $ \langle t'|\sigma_0^{\alpha}| t\rangle,$
these matrices are just the spin$-1$ matrices:
\begin{align*}
 \sigma_0^x & = \left( \begin{array}{ccc}
0 & \frac{1}{\sqrt{2}} & 0 \\
\frac{1}{\sqrt{2}} & 0 & \frac{1}{\sqrt{2}} \\
0 & \frac{1}{\sqrt{2}} & 0 \end{array} \right) , \\
\\
\sigma_0^y & = \left( \begin{array}{ccc}
0 & -\frac{i}{\sqrt{2}} & 0 \\
\frac{i}{\sqrt{2}} & 0 & -\frac{i}{\sqrt{2}} \\
0 & \frac{i}{\sqrt{2}} & 0 \end{array} \right) ,\\
\\
\sigma_0^z & = \left( \begin{array}{ccc}
1 & 0 & 0 \\
0 & 0 & 0 \\
0 & 0 & -1 \end{array} \right)
\end{align*}
and the Hamiltonian in the reduced subspace becomes 
\begin{equation}
 H_1 = J S_0^{\alpha} \sigma_{\alpha}^{\alpha}.
\end{equation}
where $\mathbf{S}_0$ represents the spin-1 at the origin.
 
Thus for ferromagnetic impurity coupling, the new terms which couple the
triplet and the rest of the Kitaev model are similar to the original Kitaev coupling and of
the same strength. We get a Kitaev-like model with a spin$-1$ at the origin and spin$-1/2$ elsewhere. 
Here the Kondo triplet does not decouple from rest of the
Kitaev model in the strong coupling limit and does not lend itself to a simple treatment, unlike the
corresponding antiferromagnetic case.

\subsection{Topological transition}

A remarkable property of the Kondo effect in Kitaev model is that
the unstable fixed point is associated with a topological transition
from the zero flux state to a finite flux state. The strong antiferromagnetic coupling
limit amounts to studying the Kitaev model with a missing site or
cutting the three bonds linking this site to the neighbors. It was shown 
in Kitaev's original paper\cite{kitaev06} that such states with an 
odd number of
cuts are associated with a finite flux, and also that these vortices
are associated with unpaired Majorana fermions and have non-abelian
statistics under exchange. It has also been shown numerically for the
gapless phase\cite{willans10} that
the ground state of the Kitaev model with one spin missing has a finite flux pinned to
the defect site. We argue the existence of a localized zero energy Majorana mode 
from the degeneracy of the ground state in presence of impurity spin and elucidate on the
nature of this zero mode.

\begin{figure}
\begin{centering}
\includegraphics[width=0.7\columnwidth]{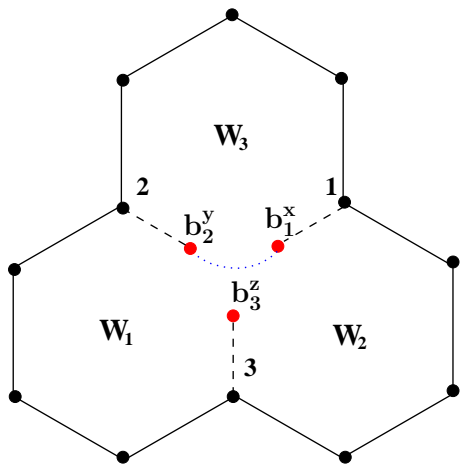} 
\par\end{centering}
\caption{\label{fig:unpaired} Schematic of the three unpaired b--Majorana fermions
formed as a result of cutting the links to the Kitaev spin at the origin. Any two of
the
three can be given an expectation value (dotted bond).}
\end{figure}

For the Hamiltonian $H=H_{0}+V_{K},$ the three plaquettes $W_{1},$
$W_{2}$ and $W_{3}$ (Fig. \ref{fig:unpaired}) that touch the impurity site are no
longer associated with conserved 
flux operators, while the flux operators that do not include the origin remain
conserved. The three plaquette operator $W_{0}=W_{1}W_{2}W_{3}$
is still conserved and $W_{0}=1$ in the ground state of the unperturbed
Kitaev model. 

We now define composite operators $\tau^{x}=W_{2}W_{3}S^{x},$
$\tau^{y}=W_{3}W_{1}S^{y}$ and $\tau^{z}=W_{1}W_{2}S^{z}$ ( $S^{\alpha}$ are the
Pauli spin matrices corresponding to the impurity). Remarkably, these composite operators
represent conserved quantities for arbitrary values of the impurity coupling. The $\tau^{\alpha}$'s
do not commute with each other and instead obey an $SU(2)$ algebra,
$[\tau^{\alpha},\tau^{\beta}]=2i\epsilon_{\alpha\beta\gamma}\tau^{\gamma}.$ This
$SU(2)$ symmetry, which is exact for all couplings is realized
in the spin-1/2 representation $\left(\left(\tau^{\alpha}\right)^{2}=1\right)$.
Clearly, all eigenstates, including the ground state are doubly degenerate
(corresponding to $\tau^{z}=\pm 1$), and this applies also to the strong coupling limit.

In the strong antiferromagnetic coupling limit $J_{K}\rightarrow\infty,$ the low
energy states will be the ones in which the spin at the origin
forms a singlet $|0\rangle$ with the impurity spin, $|\psi\rangle=|\psi
K^{-}\rangle\otimes|0\rangle.$ Here $|\psi K^{-}\rangle$ represents the low
energy states of the Kitaev model with the spin at the origin removed.
To see the action of the $SU(2)$ symmetry generators on these states,
we note that they can be written as
$\tau^{\alpha}=\tilde{W}^{\alpha}\otimes\sigma_0^{\alpha}\otimes S^{\alpha}$
and $\tilde{W}^{\alpha}$ do not involve the components of the spin
at the origin, $\sigma_0^{\alpha}.$ We then have
$\tau^{\alpha}|\psi\rangle=-(\tilde{W}^{\alpha}|\psi K^{-}\rangle)\otimes|0\rangle$.
So, in the strong coupling limit, the symmetry generators act non-trivially
only in the Kitaev model sector, implying that the low energy
states of the Kitaev model with one spin removed are all doubly degenerate,
with the double degeneracy emerging from the Kitaev sector. This is also true for the 
zero-energy mode in the single particle spectrum: the two
degenerate states correspond to the zero mode being occupied or unoccupied.

Let us examine the structure of the zero mode. Removing a Kitaev spin
creates three unpaired $b-$Majorana fermions at the neighboring sites, say,
$b_{3}^{z},$ $b_{1}^{x}$ and $b_{2}^{y}$ (Fig. \ref{fig:unpaired}).
Now $ib_{1}^{x}b_{2}^{y}$
is conserved and commutes with all the conserved flux operators $W_{i}$
but not with the two other combinations $ib_{2}^{y}b_{3}^{z}$ and
$ib_{3}^{z}b_{1}^{x}.$ So, we can choose a gauge where the expectation
value of $ib_{1}^{x}b_{2}^{y}$ is equal to +1 such that these two $b-$modes drop out of 
the problem and we equivalently have one unpaired
$b-$Majorana fermion. The unpaired $b_{3}^{z}$ Majorana has dimension
$\sqrt{2}$ and therefore, there must be an unpaired Majorana mode in
the $c-$sector (again of dimension $\sqrt{2}$) so that together these two
give the full (doubly degenerate) zero energy mode. Also, while
the $b_{3}^{z}$ mode is sharply localized, the wave function of the
$c$ mode can be spread out in the lattice.
%Existence of a zero energy $c-$Majorana mode  has also been
%shown explicitly by considering the Kitaev model with one site missing
%\cite{santhosh12}.

For the ferromagnetic case, while the strong coupling limit also leads to a model with doubly-degenerate levels, 
we are unable to explicitly identify a zero energy unpaired Majorana fermion and not address the question
as to whether a nontrivial $Z_2$ flux can be associated with closed paths enclosing the $S=1$ defect.
For this purpoose, we perform a numerical exact diagonalization analysis below.

\section{Numerical studies}\label{sec:numerical}

%Efforts to establish an experimental realisation of Kitaev model has led to 
%extensive investigations into the electronic and magnetic properties of the 
%layered iridates and osmates, where the large spin-orbit coupling effect of the
%5$d$ electrons is the essential driving force of the $J_\text{eff} = 1/2$ physics.
%The alkali iridates $A_2$IrO$_3$ ($A=$ Na, Li), seem to hold a lot of promise 
%although recent neutron and resonant x-ray scattering \cite{cao} measurements have shown
%that  Na$_2$IrO$_3$ orders into a zig-zag state below 15 K. However, microscopics
% of the system dictates that the paths   between the $Ir-O-I$r bonds between two 
%oxygen octahedra are not exactly equivalent. The deviation from the exact 90 degree
% bond angle results in the incomplete cancellation and thus this superexcahnge mechanism 
%leads to another additional Heisenberg term along with the Kitaev interaction
% resulting in the Kitaev Heisenberg (KH) model. The phase angle $\lambda$  interpolates
% the model between two limits: ($[\lambda =\pm\pi/2]$ for pure Kitaev) and
% ($[\lambda = n\pi]$ for pure Heisenberg). Although now there seems to be compelling evidence,
%that the negative KH model might \cite{khaliullin} not completely capture the 
%physics of Na$_2$IrO$_3$, nevertheless it does have a significant region in the phase space
%which is occupied by the zig-zag ground state. For other values of $\lambda$, the stripy, 
%N\'eel, ferromagnetic and spin liquid ground states are obtained. The boundaries between these
%different phases is obtained by exact diagonalisation. 
%
We have used a modified Lanczos algorithm to calculate the ground state properties of a finite Kitaev 
fragment exchange-coupled to an external $S=1/2$ impurity spin as discussed in the previous sections. 

\begin{figure}
\includegraphics[width=0.9\linewidth]{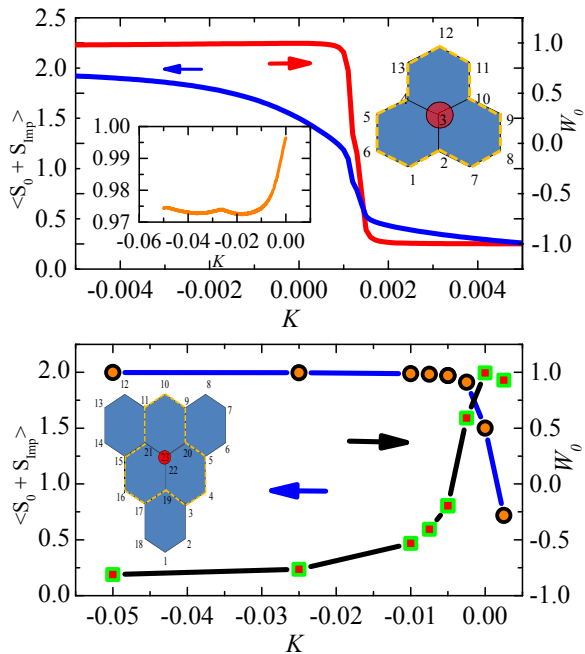} 
\caption{(Color online) {\it{(Top)}} Evolution of expectation value of the flux operator ($W_0$) and the 
total spin at the impurity site with $K$ evaluated on a three hexagon fragment. The impurity spin is coupled to 
the central spin and the flux is calculated over the boundary. The inset shows the slight decrease of $W_0$ from 1 for 
ferromagnetic coupling whereas for antiferromagnetic coupling $W_0$ changes to -1. {\it{(Bottom)}} $W_0$ and total spin at 
the impurity site evaluated for a six hexagon fragment showing the transition to $W_0$ = -1 state for ferromagnetic coupling.
}
\label{fig_imp_flux}
\end{figure}

For the antiferromagnetic case, where we already know that the strong-coupling limit $K/J\rightarrow\infty$ 
is associated with a nontrivial flux $W_0 = -1$ at the defect site, an
exact-diagonalization calculation with a fragment as small as three hexagons (open boundary conditions, impurity spin coupled to central site)
is sufficient to confirm $W_0 = -1.$ Figure~\ref{fig_imp_flux} shows the expectation value of the flux operator 
$W$ and total spin $S_{Tot} = S + s_i$ as a function of the impurity coupling $K.$  It is seen that the $W_0 = 1$ for $K=0$ 
(i.e in the pure Kitaev case) as it should be, since the ground state of Kitaev model is flux free, whereas it changes to -1 as $K$ is increased, implying 
a finite $Z_2$ flux at the origin. Within numerical accuracy, it also appears that the topological transition from $W_0 = 1$ to $W_0 = -1$ practically
occurs at the same value of $K$ at which a bound singlet state is formed between the impurity spin and the Kitaev host. For negative values of $K$ for this
three-hexagon fragment, $W_0$ appears to stay close to one implying a flux-free state even as the total spin at the defect site begins approaching $S=1.$
We suspect this anomalous result is not generally true for larger fragments and may have originated from the presence of a large number of boundary spins
that interact with the central spin. We thus performed exact-diagonalization calculations with a larger fragment with
six hexagons (open boundary conditions, impurity spin coupled to central site). As we increase the ferromagnetic impurity coupling, we clearly observe 
the $W_0 =1$ to $W_0 = -1$ topological transition. Once again, the topological phase transition and the magnetic transition (at which the total spin at the defect
site becomes $S=1,$ are practically coincident.

\section{Two-channel Kondo behavior}\label{sec:two-channel}
In Ref.~\onlinecite{willans10}, the temperature and magnetic field dependences of the magnetic susceptibility of the Kitaev model 
with a missing site were obtained as $\chi_{\text{imp}}(T) \sim \ln(1/T)$ and $\chi_{\text{imp}}(h)\sim\ln(1/h),$ respectively,
which bears striking resemblance 
to the low-temperature impurity susceptibility in the two-channel Kondo model~\cite{emery,sengupta}. This is not a mere coincidence.
In the two-channel Kondo problem, it is long known~\cite{coleman1995} that the low energy physics is described by a model of
a localized, zero energy Majorana fermion interacting with a band of dispersing Majorana fermions with
a finite density of states at the Fermi energy. Likewise, in the Kitaev model with a missing site, we can
clearly identify a localized zero energy Majorana $b-$fermion coexisting with dispersing Majorana $c-$fermions
with a \emph{nonvanishing} density of states~\cite{willans10} at zero energy. A nonvanishing density of states for the dispersing
Majoranas is a very unusual result for a honeycomb lattice, and is associated with the fact that the
missing site is associated with a finite $Z_2$ flux. If one instead estimates the density of states
for the dispersing fermions in the absence of a $\pi-$flux, the density of states would vanish~\cite{willans10} at the Fermi energy; indeed, this
would be the case in graphene. To compute the magnetic susceptibility, we choose 
a gauge where $b_3^z$ is the zero energy localized Majorana fermion. Using $S_3^z = i c_3 b_3^z$ 
the magnetic susceptibility of the defect can be expressed as
\begin{align}
 \chi_{\text{imp}} = T\sum_{n,k}\frac{1}{i\nu_n - \epsilon_k}\frac{1}{i\nu_n} \sim \ln(1/T),
 \label{chi_imp1}
\end{align}
where $(i\nu_n - \epsilon_k)^{-1}$ and $1/i\nu_n$ are respectively the Green functions
of the dispersing and the (zero-energy) localized Majorana fermions and we used the fact that
the density of states of the dispersing Majorana fermions does not vanish at zero energy. For finite fields $h$
in the low temperature limit, the logarithmic divergence of Eq.\ref{chi_imp1} gets cut off by the field, and 
one obtains $\chi_{\text{imp}}(h) \sim \ln(1/h).$ For the ground state entropy, one notes that the 
dispersing Majorana fermions have zero entropy at $T=0$ while the localized zero energy Majorana
fermion has a finite entropy $S = (1/2)\ln(2).$ This again agrees with the two-channel Kondo result.\cite{emery}

\section{Discussion}\label{sec:discussion}
In summary, we showed that the problem of spinless and spinful defects in the honeycomb $S=1/2$ Kitaev model can be approached
from a more general ``Kondo perspective'' of local exchange coupling of external paramagnetic impurities with
a host spin. On one hand, such an approach gives us a new class of Kondo effects where the magnetic binding-unbinding
transition is accompanied by a change of topology of the ground state. On the other hand, some intriguing recent
observations, such as logarithmic singularities in the magnetic response of Kitaev models with vacancies,
are now recognizable as familiar Kondo stories - in this case, we note a remarkable similarity with the two-channel Kondo problem.
It would be interesting to study a lattice of vacancies in the Kitaev model from the perspective of a two-channel Kondo lattice.
One would like to better understand the Kitaev model with a $S=1$ defect. This nonintegrable nature of this problem prevents us
from repeating the kind of analysis one could make for the vacancy case where similarity with the two-channel Kondo problem was
observed. Our numerical approach, based on exact diagonalization calculations of relatively small fragments, cannot answer questions such 
as the density of states of low energy excitations. Another direction for future study would be to consider a more general
Kitaev-Heisenberg model and track the Kondo effect as a function of the relative strengths of Kitaev and Kondo interactions.
This should give insights into magnetic impurity response in Kitaev candidate materials where Kitaev and Heisenberg
interactions are believed to compete with each other.

\begin{acknowledgments}
 We are grateful for useful discussions with K. Damle.
 V.T. acknowledges financial support from Argonne Natl. Lab., the University of Chicago Center in Delhi, and DST (India) Swarnajayanti 
grant (no. DST/SJF/PSA-0212012-13). S.D.D acknowledges the financial support provided by Cambridge Commonwealth Trust(CCT) and hospitality 
provided by DTP (TIFR).
\end{acknowledgments}

\renewcommand{\bibname}{References}

\end{document}